\documentclass{article}

\usepackage[version=4]{mhchem}
\usepackage{enumitem}
\usepackage{siunitx} 
\usepackage{bm}
\usepackage{caption}
\usepackage{array}
\usepackage{xcolor}

\usepackage{amssymb}
\usepackage{upgreek}

\usepackage{arxiv}

\usepackage[utf8]{inputenc} 
\usepackage[T1]{fontenc}    
\usepackage{hyperref}       
\usepackage{url}            
\usepackage{booktabs}       
\usepackage{amsfonts}       
\usepackage{nicefrac}       
\usepackage{microtype}      
\usepackage{cleveref}       
\usepackage{lipsum}         
\usepackage{graphicx}
\usepackage{doi}

\usepackage[numbers,sort&compress]{natbib}

\usepackage{graphicx}
\usepackage{color}
\usepackage{floatrow}
\usepackage{subfig}
\usepackage{mwe}
\usepackage{caption}
\usepackage{afterpage}
\usepackage{changepage}
\usepackage{hyperref}
\hypersetup{
     colorlinks   = true,
     citecolor    = blue
}
\usepackage[version=4]{mhchem}
\usepackage{adjustbox}

\usepackage{ulem}

\usepackage{amssymb}
\usepackage{amsthm}
\usepackage{amsmath,bm}
\allowdisplaybreaks[4]
\usepackage{bibentry}
\usepackage{mathrsfs}
\usepackage[T1]{fontenc}  
\usepackage[labelsep=period,labelfont=bf,figurename=Fig.]{caption}
\usepackage{tikz}
\usepackage{pgfplots}
\usetikzlibrary{arrows,shapes,trees}
\usetikzlibrary{decorations.pathreplacing}
\usetikzlibrary{intersections,backgrounds}
\usetikzlibrary{calc}
\usetikzlibrary{positioning}
\usepackage{tkz-euclide}
\tikzset{>=latex}
\usepackage{colortbl}

\usepackage{pict2e}
\usepackage{overpic}
\usepackage{multirow}
\usepackage{calc}
\usepackage{longtable}
\usepackage{pbox}
\usepackage{array}
\newcolumntype{L}[1]{>{\raggedright\let\newline\\\arraybackslash\hspace{0pt}}m{#1}}
\newcolumntype{C}[1]{>{\centering\let\newline\\\arraybackslash\hspace{0pt}}m{#1}}
\newcolumntype{R}[1]{>{\raggedleft\let\newline\\\arraybackslash\hspace{0pt}}m{#1}}

\title{Assessment of turbulent fire dynamics and combustion instabilities using a flamelet model}

\newif\ifuniqueAffiliation
\uniqueAffiliationtrue

\ifuniqueAffiliation 
\author{ \href{https://orcid.org/0000-0003-3133-4292}{\includegraphics[scale=0.06]{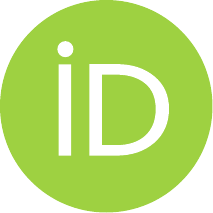}\hspace{1mm}Yunxiao Yan} \\
	University of Science and Technology of China \\
    State Key Laboratory of Fire Science \\
    Huangshan Road 443, 230026 Hefei, PR China 
	\And
	\href{https://orcid.org/0000-0002-5404-0855}{\includegraphics[scale=0.06]{orcid.pdf}\hspace{1mm}Fabian Brännström} \\
    University of Wuppertal \\
    Chair of Fire Dynamics \\
    Gau{\ss}stra{\ss}e 20, Wuppertal, 42119, Germany
	\And
	\href{https://orcid.org/0000-0001-9333-0911}{\includegraphics[scale=0.06]{orcid.pdf}\hspace{1mm}Christian Hasse} \\
    Technical University of Darmstadt\\
    Simulation of Reactive Thermo-Fluid Systems\\
    Otto-Berndt-Str. 2, 64287 Darmstadt, Germany
	\And
	\href{https://orcid.org/0000-0003-3133-4292}{\includegraphics[scale=0.06]{orcid.pdf}\hspace{1mm}Xu Wen} 
    \thanks{Corresponding author: xu.wen@ustc.edu.cn} \\	
	University of Science and Technology of China \\
    State Key Laboratory of Fire Science \\
    Huangshan Road 443, 230026 Hefei, PR China 
    }

\begin{document}

\maketitle

\begin{abstract}
The Sandia one-meter methane fire plume is an established benchmark for turbulent combustion modeling of large-scale flames. This study investigates the combustion instabilities formed close to the base of the Sandia fire plume. Finite rate chemistry and differential diffusion are considered using a flamelet/progress variable (FPV) approach. The performance of the FPV approach is assessed by comparing with the eddy dissipation model (EDM) and the experimental data for the large-scale fire plume via large eddy simulations (LES). The effects of radiation modeling and mesh resolution on the predictive capability of the model are systematically investigated by comparing the axial and radical velocities against the experimental data at various locations. Although all models successfully capture the primary flow characteristics of fire plumes, the FPV model with differential diffusion yields improved predictions in the near-flame-base region. The formation mechanism of cellular flow structures near the flame base is investigated via a \textit{budget analysis} of the vorticity equation, and the type of instability governing the formation of the cellular structure is clarified. Finally, the individual effects of finite rate chemistry and differential diffusion on the prediction of the thermo-chemical quantities are quantified. Overall, this study explains the underlying physics governing combustion instabilities at the base of turbulent fire plumes, provides novel insights into the performance of flamelet models for LES of gaseous pool fires, and offers reliable guidance for the high-fidelity numerical simulation of large-scale turbulent buoyancy driven flames.
\end{abstract}


\section*{Highlights}

\begin{itemize}
\item Flamelet/LES is applied to predict the combustion instabilities in buoyancy driven flames with differential diffusion;
\item The mechanism governing the combustion instability characterized by periodic cellular structures near the flame base is analyzed and quantified for the first time;
\item The effects of finite rate chemistry and differential diffusion on LES of pool fire combustion are quantified.
\end{itemize}


\section{Introduction}
\label{Sec:1}

As highlighted in the recent review by Merci \cite{merci_modelling_2026}, computational fluid dynamics (CFD) simulations serve as a valuable tool for exploring buoyancy-driven fires. Previous CFD simulations mainly focus on the prediction of overall fire dynamics, and only several studies \cite{domino2021predicting, taha2024large, desjardin2004large} investigated the combustion instabilities in fire plumes. The instabilities in turbulent buoyancy driven flames are important phenomena governing the overall fire dynamics, and the mass and heat transfer in the combustion process. In particular, a systematic analysis of the combustion and flame instabilities at the flame base featuring streaks in flame fronts has not been reported.

\begin{figure}[!h]
    \centering
    \captionsetup[subfigure]{labelformat=empty} 
    \subfloat[]{
    \begin{minipage}[b]{0.6\textwidth}
\tikz[remember picture] \node[inner sep=0pt,outer sep=0pt] (a) 
{\includegraphics[trim = 0mm 0mm 0mm 0mm, clip, angle=0, width=1.\linewidth]{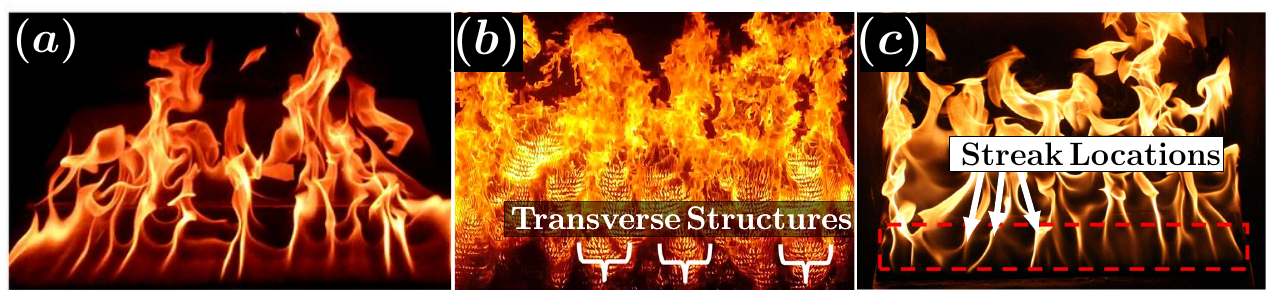}};
\end{minipage}
\begin{tikzpicture}[remember picture,overlay]
\end{tikzpicture}}
    \vspace{-5mm}
    \caption{\label{fig:exp_cellular} Snapshots of flame structures from different experiments illustrating combustion instabilities and streamwise streaks in flame base. (a) Top view of flames from an inclined ethylene burner \cite{gorham2014studying}; (b) A wind-tunnel experimental fire spreading in cardboard fuel \cite{finney2015role}; (c) The sample top-view image of heptane fire \cite{miller2015investigation}. Brackets in (b) indicate the buoyant instabilities forming as transverse structures. Streak locations are indicated in (c). }
\end{figure}

Combustion instabilities in fire plumes have been widely investigated through experimental methodologies. For example, Cetegen et al.~\cite{cetegen2000experiments} identified two primary instability modes in buoyant diffusion flames, i.e., the sinuous meandering mode and the varicose mode. They found that flames may switch between the two combustion modes. Hu et al.~\cite{hu2015flame} further refined this classification for ethanol pool fires and identified three additional distinct modes, i.e., short-life Rayleigh-Taylor (R-T) instabilities, extended R-T instabilities, and puffing instabilities. They demonstrated that the dominant instability mode can be different for different pool sizes or lip heights. Hu \cite{hu2017review} summarized the different modes of combustion instability in small-to-medium pool fires, including sinuous meandering, varicose mode, R-T instabilities, extended R-T instabilities, and puffing instabilities. While these instability modes have been well-established, recent experimental studies \citep{gorham2014studying, finney2015role, miller2015investigation, chen2023pool, sung2024global} have revealed another type of coherent structure: streamwise streaks in flame base. Gorham et al.~\citep{gorham2014studying} identified the existence of streamwise streaks as flame peaks and troughs (as shown in Fig.~\ref{fig:exp_cellular}a), attributing their formation to a centrifugal instability generated by the interaction between incoming boundary layer vorticity and flame vorticity. Finney et al.~\citep{finney2015role} observed streamwise streak structures in flame base of a wind-tunnel fire spreading over cardboard fuel (as shown in Fig.~\ref{fig:exp_cellular}b). They attributed these streaks to counter-rotating Taylor-Görtler vortex pairs, whose convergence zones generate alternating upward and downward flows that periodically force flames and hot gases downward and forward, causing them to splay horizontally into the characteristic flame ``towers and troughs''. Miller et al.~\cite{miller2015investigation} also found coherent streak structures (see Fig.~\ref{fig:exp_cellular}c) and attributed them to Rayleigh-Taylor instabilities. Furthermore, they reported that the burner depth significantly affects the number of streaks. Due to the limited experimental data, the reason for the formation of combustion instabilities has not been clarified. For the Sandia one-meter diameter fire plume, which is a target case in the MaCFP workshop \cite{brown2018proceedings}, the combustion instabilities that are observed at the flame base have not been investigated.

The phenomena of puffing mode instabilities were also reported in previous works through numerical simulations \cite{desjardin2004large, blanquart2008large, ma2020large, domino2021predicting, taha2024large}. For example, Ma et al.~\cite{ma2021exploring} reported that the natural laminar instability near the plume edge cannot be captured when the grid size is coarser than 1\,cm. The streamwise streaks in flame base showing the ``finger-like'' azimuthal instabilities were captured by DesJardin et al.~\cite{desjardin2004large} and Taha et al.~\citep{taha2024large}. For the Sandia one-meter diameter helium plume, Taha et al.~\cite{taha2024large} concluded that finger-like structures are formed due to the combined effects of gravity and baroclinic vorticity. Similarly, DesJardin et al.~\cite{desjardin2004large} think the misalignment of density gradient and gravity acceleration vectors is the dominant factor. Finger-like flame structures were also reported by Domino et al.~\cite{domino2021predicting} at the flame base of a five-meter diameter heptane pool fire using an unsteady flamelet approach. The instability is attributed to the interaction between radial entrainment toward the fire center and Rayleigh-Taylor instabilities. However, the underlying physics governing these combustion instabilities in large-scale fire plumes, and the role of different types of instabilities are not clear.

For combustion modeling of buoyancy-driven fires, infinitely fast chemistry models are widely used, such as the Eddy Break Up (EBU) model \citep{spalding1971mixing}, the Eddy Dissipation Model (EDM) \citep{magnussen1977mathematical} and the Eddy Dissipation Concept (EDC) model \citep{magnussen1981structure}, which can provide an overall prediction of flame dynamics \cite{wang2011large, maragkos2012application, chen2014large, maragkos2017large, ren2019modeling, maragkos2020use, lapointe2020efficient, maragkos2022analysis, maragkos2023sub, maragkos2023flame, merci2023importance, snegirev2023development, taha2024large}. Although infinitely fast chemistry models are widely adopted for fire simulations owing to their high computational efficiency and simplicity, flamelet \citep{peters1984laminar} is one of the very popular models for turbulent flames - from laboratory configurations to technical systems such as gas turbines. Infinitely fast chemistry models assume that turbulent mixing is rate-limiting and are therefore unable to accurately capture the effects of finite-rate chemical kinetics. Thus, finite-rate chemistry models are preferred to reproduce non-equilibrium behaviors such as local extinction and ignition. As pointed out by Merci \cite{merci_modelling_2026}, the application of the flamelet model to enclosure fires requires special attention. On the one hand, the generation of the flamelet table requires the exact fuel compositions and the corresponding detailed chemical mechanisms, which are hard to determine in real fires. On the other hand, the heat release rate (HRR), which characterizes the overall fire dynamics such as flame height \cite{auth2019using}, cannot be directly obtained from the flamelet model. Nevertheless, for fires with well-defined fuels, the flamelet model is promising since it can incorporate detailed chemical mechanisms without high computational costs. In fact, several research attempts have been made to apply flamelet models in different formulations to high-fidelity fire simulations \citep{cheung2009fully, han2021pdf, domino2021predicting, nmira2023large, lin2023numerical, lin2024radiative}. For the Sandia one-meter diameter methane plume, both the infinitely fast chemistry model \cite{taha2024large,maragkos2017large} and the flamelet model \cite{cheung2009fully, han2021pdf} have been employed to predict its combustion characteristics. However, the combustion instabilities are not captured or not analyzed in these studies. In addition, a side by side comparison, the analysis and the critical assessment of both infinitely fast and finite chemistry models are also lacking in the literature.

As fire simulation involves the transition from laminar to turbulent flow and differential diffusion (DD) is expected to play a vital role in laminar flows, it is essential to consider the effects of differential diffusion in fire simulations. Differential diffusion is a phenomenon that appears due to the differences in species' thermal diffusivity and its mass diffusivity, while preferential diffusion, i.e., non-equal Lewis number of species, describes the differences in species' mass diffusivity. Although the laminar flow may develop into turbulence within a limited distance, the effects of differential diffusion may persist and build up in the downstream region \cite{barlow2015local, pitsch1998unsteady, tang2023differential}. Differential diffusion is a phenomenon that appears due to differences in species' diffusivities. On the basis of the EDC model, Maragkos et al.~\cite{MARAGKOS201722, maragkos2019towards} investigated the effects of differential diffusion on two different methanol gaseous pool fires based on a one-step infinitely fast chemistry assumption, and concluded that differential diffusion is not important for the specific case. However, Wu et al.~\citep{wu2021limitations} reported that preferential diffusion plays an important role in the dynamics near the pool fire due to the nearly laminar behavior through a \textit{budget analysis} and the inclusion of detailed chemistry in the simulations. They also stated that the effects of preferential diffusion may persist and even further amplify in large pool fires, such as the Sandia one-meter diameter pool fire. In this work, the role of differential diffusion in large-scale fire plumes will be clarified through high-fidelity flamelet/LES simulations.

The purpose of this work is to \textcolor{black}{apply the flamelet model with differential diffusion to predict the combustion instabilities in the Sandia one-meter methane fire plume. The characteristics and formation mechanisms of coherent streaks associated with combustion instabilities at the flame base are systematically analyzed, and the contributions from different processes are quantified based on a \textit{budget analysis} of the vorticity equation.} The effects of radiation models on the prediction accuracy are evaluated by comparing different radiative property models that are widely used in fire simulations. The effects of finite rate chemistry are assessed by comparing the simulation results obtained from the infinitely fast chemistry combustion model EDM and the finite rate chemistry combustion model flamelet based on the unity Lewis number assumption. Finally, the assumption of unity Lewis number that is often adopted in fire simulations is examined by flamelet/LES of large-scale pool fires with and without differential diffusion being incorporated.

\section{Governing equations}
\label{Sec:2}
In the context of LES, the governing equations for mass and momentum are solved in the Favre-filtered formulations for both the EDM and flamelet models. The sub-grid scale (SGS) viscosity is calculated with the $k$-equation turbulence model \cite{yoshizawa1986statistical}.
\subsection{EDM model} 
\label{Subsec:21}

In the EDM model \citep{magnussen1977mathematical}, the fuel combustion rate is assumed to be infinitely fast. Thus, the reaction is expressed as $\mathrm{CH_4 + O_2 \rightarrow CO_2 + H_2O }$ for methane combustion. The Favre-filtered governing equations for the enthalpy and major species mass fractions, i.e., $\widetilde{ Y_{\rm CH_4} }$, $\widetilde{ Y_{\rm O_2} }$, $\widetilde{ Y_{\rm CO_2} }$, and $\widetilde{ Y_{\rm H_2O} }$, are solved. The EDM model is formulated based on the unity Lewis number assumption, referred to as ``$\rm EDM\_Le1$''. In the $\rm EDM\_Le1$ model, the chemical reaction rate for species $i$, denoted as $\dot{\omega}_i$, is calculated by relating it with the fuel reaction rate, i.e., $\dot{\omega}_i = s_i  \dot{\omega}_{\ce{CH4}}$. Here, $s_i$ is the stoichiometric coefficient of species $i$. The chemical reaction rate of \ce{CH4} $\dot{\omega}_{\ce{CH4}}$ is calculated as, 
\begin{equation}\label{eq:wp1}
\overline{\dot{\omega}_{\ce{CH4}}} = \overline{\rho} C_{\mathrm{EDM}} \frac{\widetilde{\epsilon}}{\widetilde{k}} \rm {min}\left( \widetilde{Y_{\ce{CH4}}}, \dfrac{\widetilde{Y_{\mathrm{O_2}}} }{s_{\ce{O2}}} \right) 
\end{equation}
where $\left( \overline{\cdot} \right)$ and $\left( \widetilde{\cdot} \right)$ denote the spatially-averaged and Favre-filtered variables, respectively. $\rho$ is the gas density, $C_{\mathrm{EDM}}$ is the modeling constant, which is set to 1, $k$ the turbulence kinetic energy, $\epsilon$ the eddy dissipation rate, $Y_{\ce{CH4}}$ and $Y_{\mathrm{O_2}}$ the mass fractions of fuel and \ce{O2}, respectively.

\subsection{Flamelet model}
\label{Subsec:22}

In the flamelet model \citep{peters1984laminar}, the governing equations of several trajectory variables are solved instead of those for species mass fractions and enthalpy. For the fire flame studied, the mixture fraction $Z$ is introduced to characterize the mixing process, which is defined based on the elemental mixture fractions by Bilger \citep{bilger1990reduced} to consider the effects of differential diffusion. 
The progress variable $C$ is introduced to characterize the progress of reactions \citep{pierce2001progress}, which is defined as, $C = Y_{\mathrm{CO_2}} + Y_{\mathrm{H_2O}} + Y_{\mathrm{CO}} + Y_{\mathrm{H_2}} $ \citep{ihme2008modeling}. The total enthalpy $H_e$ is introduced to  characterize the heat losses in fire simulation. To facilitate flamelet table access, $H_e$ is normalized as $H_{e,\rm norm} = \left(H_e - H_{e,\min} \right) / \left( H_{e,\max} - H_{e,\min} \right)$, where $H_{e, \max}$ and $H_{e, \min}$ are the maximum and minimum values of $H_e$ for specific values of $Z$ and $C$, respectively. In this work, two flamelet tables are generated with the mass diffusion either calculated with the unity Lewis number assumption (referred to as ``$\rm FPV\_Le1$'') or the mixture-averaged approach (referred to as ``$\rm FPV\_DD$''). The GRI-Mech 3.0 chemical mechanism \citep{smith2011gri} is utilized to characterize the chemistry. The flamelet solutions are tabulated as a function of $Z$, $C$ and $H_{e,\rm norm}$ as $\Psi = \Psi \left(Z, C, H_{e,\rm norm} \right)$. The turbulence-chemistry interactions are considered with the presumed probability density function. In this work, a $\delta$-PDF is used to describe the distributions of $C$ and $H_{e,\rm norm}$, and a $\beta$-PDF is used to characterize the distributions of $Z$. The $\beta$-PDF requires the variance of $Z$, i.e., $\widetilde{Z''^2}$, to be introduced. Thus, the Favre-filtered flamelet solutions are tabulated as, $ \widetilde{ \Psi } = \Psi \left(\widetilde{Z}, \widetilde{C}, \widetilde{H_{\rm e,norm}}, \widetilde{Z''^2} \right)$

The Favre-filtered transport equations for the trajectory variables $\widetilde{Z}$, $\widetilde{C}$ and $\widetilde{H_{e}}$ are solved in the flow solver \citep{wen2021flamelet, wen2025flamelet},
\begin{equation}\label{eq:FPVZ}
\frac{\partial\left(\bar{\rho}\widetilde{Z}\right)}{\partial t} +\nabla\cdot{\left(\bar{\rho}\widetilde{\bm{ u }}\widetilde{Z}\right)} =\nabla\cdot\Bigg[\left(\frac{\bar{\lambda}}{\bar{c}_p}+\bar{\rho}D_t\right)\nabla{\widetilde{Z}}\Bigg]
+\underbrace{\nabla\cdot\Bigg(\mathfrak{D}_{Z}\nabla{\widetilde{C}}\Bigg)}_{\Lambda_{Z}^{\mathrm{DD}}}
\end{equation}
\begin{equation}\label{eq:FPVC}
\frac{\partial\left(\bar{\rho}\widetilde{C}\right)}{\partial t} +\nabla\cdot{\left(\bar{\rho}\widetilde{\bm{ u }}\widetilde{C}\right)} 
=\nabla\cdot\left[\left(\frac{\bar{\lambda}}{\bar{c}_p}+\bar{\rho}D_t\right)\nabla{\widetilde{C}}\right]
+\underbrace{\nabla\cdot\left(\mathfrak{D}_{C}\nabla{\widetilde{C}}\right)}_{\Lambda_{C}^{\mathrm{DD}}}+\overline{\dot{\omega} }_{C}
\end{equation}
\begin{equation}\label{eq:FPVHe}
\frac{\partial\left(\bar{\rho}\widetilde{H_e}\right)}{\partial t} +\nabla\cdot{\left(\bar{\rho}\widetilde{\bm{ u }}\widetilde{H_e}\right)} =\nabla\cdot\Bigg[(\bar{\rho} \widetilde{\alpha}+\bar{\rho}\alpha_t)\nabla{\widetilde{H_e}}\Bigg]
+\underbrace{\nabla\cdot\Bigg(\mathfrak{D}_{H_e}\nabla{\widetilde{C}}\Bigg)}_{\Lambda_{H_e}^{\mathrm{DD}}}+\overline{\dot{S}_{rad}}
\end{equation}
where $t$ is the Euler-like time, $\bm{ u }$ the velocity vector, $\lambda $ and $c_p$ are the thermal conductivity and specific heat capacity of the mixture, respectively. $D_{\rm t}$ is the sub-grid scale mass diffusivity, $ \alpha $ the thermal diffusivity, $\alpha_t$ the sub-grid scale thermal diffusivity. ${\dot{\omega} }_{C}$ is the chemical reaction rate of $C$. The radiation term ${\dot{S}_{rad}}$ is calculated using the finite volume discrete ordinate method (fvDOM) \citep{chai1994finite}. The gas absorption coefficients are calculated using the gray-mean absorption emission (GMAE) model, with the corresponding coefficients taken from Ref.~\cite{barlow2001scalar}. The results are compared with those obtained using the weighted sum of gray gases (WSGG) model \citep{modest1991weighted} to assess the influence of the radiation model on the prediction accuracy. The WSGG model has been widely used in gas-radiation calculations \cite{ren2025cfd}, with extensions to account for variations in the \ce{H2O}/\ce{CO2} concentration ratio \cite{johansson2011account} and to refit the absorption coefficients based on the HITEMP 2010 spectral emissivity database \cite{bordbar2014line}. The $\Lambda_{({\cdot)}}^{\mathrm{DD}}$ term in the above equations quantifies differential diffusion for the variable $({\cdot)}$. The differential diffusion coefficients $\mathfrak{D}_{Z}$, $\mathfrak{D}_{C}$ and $\mathfrak{D}_{H_e}$ are calculated as \citep{donini2015differential, de2010inclusion},
\begin{equation}\label{eq:psi1}
\mathfrak{D}_{Z}=\frac{\bar{\lambda}}{\bar{c}_{p}}\sum_{i=1}^{N_{s}}\zeta_{i}\Big(\frac{1}{Le_{i}}-1\Big) \Gamma_i
\end{equation}
\vspace{-0.5mm}
\begin{equation}\label{eq:psi2}
\mathfrak{D}_{C}=\frac{\bar{\lambda}}{\bar{c}_{p}}\sum_{i=1}^{N_{s}}\varsigma_{i}\Big(\frac{1}{Le_{i}}-1\Big) \Gamma_i
\end{equation}
\vspace{-0.5mm}
\begin{equation}\label{eq:psi3}
\mathfrak{D}_{H_e}=\frac{\bar{\lambda}}{\bar{c}_{p}}\sum_{i=1}^{N_{s}}h_i\Big(\frac{1}{Le_{i}}-1\Big) \Gamma_i
\end{equation}
where $\zeta_{i}$ is the coefficient for species $i$ in the Bilger mixture fraction, $\varsigma_{i}$ the coefficient for species $i$ in the definition of $C$, and $h_i$ the total enthalpy of species $i$. $N_s$ is the number of species in the chemical reaction mechanism, and $Le_i$ the Lewis number of species $i$. The term $\Gamma_i$ quantifies the contribution of species $i$ to the diffusion flux, which is estimated as,
\begin{equation}\label{eq:psi4}
\Gamma_i = \frac{\partial Y_{i}}{\partial Z}+\frac{\partial Y_{i}}{\partial C}\frac{\partial C^{\rm 1D}}{\partial Z} +\frac{\partial Y_{i}}{\partial H_{e,norm}}\frac{\partial H_{e,norm}^{\rm 1D}}{\partial Z}
\end{equation}
where the superscripts $1D$ in $C^{\rm 1D}$ and $H_{e,norm}^{\rm 1D}$ indicate that $C$ and $ H_{e,\rm norm} $ are functions of $Z$ only in the 1D flamelet solutions following previous works \citep{wen2021flamelet, donini2015differential}. The above terms are calculated based on the flamelet solutions and stored in the flamelet table. The formulation for the governing equation of $\widetilde{Z''^2}$ is referred to in Pera et al.~\citep{pera2006modeling}. The flamelet model with similar formulations has been validated in previous works for gaseous combustion \citep{wen2021flamelet, donini2015differential} and solid fuel combustion \citep{wen2025flamelet}. 

The flamelet model based on the unity Lewis number assumption, i.e., the ``$\rm FPV\_Le1$'' model, can be recovered when the terms $\Lambda_{({\cdot)}}^{\mathrm{DD}}$ in Eqs.~\eqref{eq:FPVZ}-\eqref{eq:FPVHe} are neglected. 

\section{Numerical setup}
\label{Sec:3}

\begin{figure}[!h]
    \centering
    \captionsetup[subfigure]{labelformat=empty} 
    \subfloat[]{
    \begin{minipage}[b]{0.5\textwidth}
\tikz[remember picture] \node[inner sep=0pt,outer sep=0pt] (a) 
{\includegraphics[trim = 0mm 0mm 0mm 0mm, clip, angle=0, width=1.\linewidth]{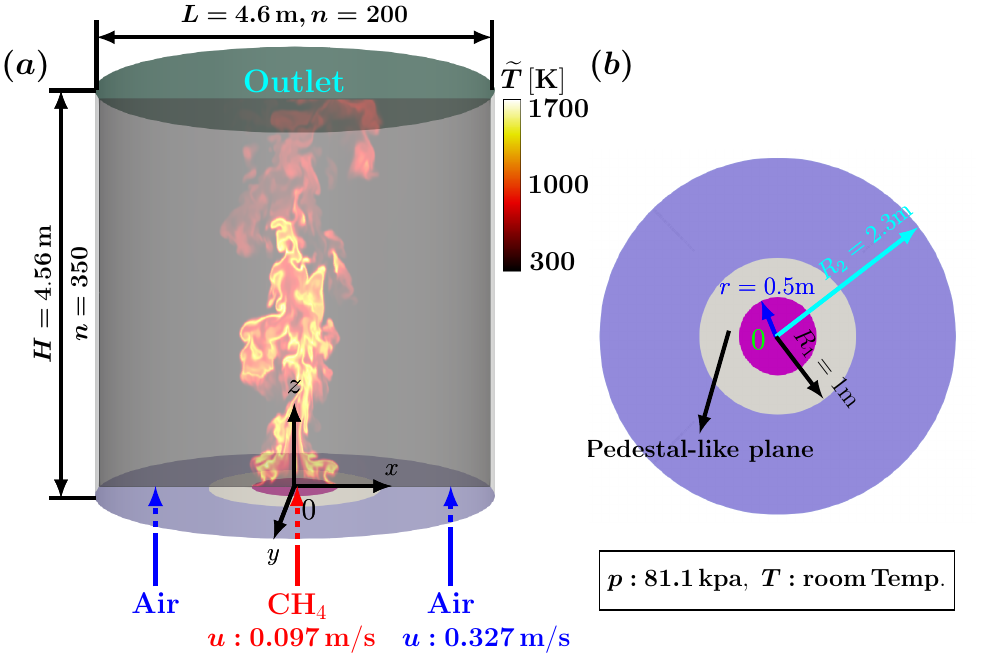}};
\end{minipage}
\begin{tikzpicture}[remember picture,overlay]
\end{tikzpicture}}
    \vspace{-8mm}
    \caption{\label{fig:boundary} Computational setup for the Sandia one-meter diameter methane fire plume. The boundary conditions and the grid setup are superimposed. (a) Front-view showing the temperature distribution in the $x$-$z$ plane for the $\rm FPV\_DD$ case. (b) Bottom-view highlighting the inlet dimensions. }
\end{figure}

The numerical models are evaluated by simulating the established benchmark of the Sandia one-meter diameter fire plume \citep{tieszen2002experimental}. As schematically shown in Fig.~\ref{fig:boundary}, the methane fuel enters the domain through a 1\,m diameter inlet at room temperature and a constant velocity of 0.097\,m/s \citep{tieszen2002experimental}. The fuel inlet is surrounded by a pedestal-like plane with a radius of 1\,m. The velocity in the coflow is set to be constant and equal to 0.327\,m/s. The setup of the boundary conditions is the same as in previous studies \citep{han2021pdf, lin2024radiative}. In the EDM model, the species mass fractions of fuel and oxidizer at the inlets are specified as Dirichlet boundary conditions, corresponding to pure methane and air, respectively. In the flamelet simulation, the inlet boundary conditions for the trajectory variables of $\widetilde{ Z } $, $\widetilde{ Z''^2}$, $\widetilde{C}$ and $\widetilde{H_{e}}$ are set as follows. The values of $\widetilde{Z}$, $\widetilde{Z''^2}$, $\widetilde{C}$ and $\widetilde{H_{e}}$ at the central inlet are equal to 1, 0, 0, and $-1.27 \times 10^6 \, \mathrm{J/kg}$, respectively, while they are set to 0, 0, 0, and $1907 \, \mathrm{J/kg}$ at the coflow. For the pressure, a hydrostatic initial distribution is set with a reference pressure $p_{ref}$ of 81.1 kPa, i.e., ${p} = p_{ref} + p_{rgh} +  {\rho} \cdot g \cdot h $, where the hydrostatic pressure $p_{rgh}$ is set to 0 at the inlets and the height is calculated by setting the flame base as the reference location. The temperature distribution in the $x$-$z$ plane is shown in Fig.~\ref{fig:boundary}a for the $\rm FPV\_DD$ case, which clearly illustrates that the gaseous fuel ignites close to the source, leading to a rapid increase in temperature. A detailed analysis of this flame is presented in Section~\ref{Subsec:42}.

In the computational setup, the computational domain features a cylinder with a height of 4.56\,m and a diameter of 4.6\,m. Two mesh resolutions with 22.4\,M and 14.9\,M grid points are adopted to evaluate the quality of the LES results. Specifically, for the finer mesh, the computational domain is discretized with 350, 200 and 80 grid points in the axial, radial and circumferential directions, respectively, with local refinement to improve spatial accuracy. The minimum and maximum cell sizes are 7.6\,mm and 21.9\,mm, respectively. For the coarse mesh, the domain is divided into 300, 180, and 70 grid points in the axial, radial and circumferential directions, respectively, with minimum and maximum cell sizes of 8.7\,mm and 23.7\,mm, respectively. The adopted mesh resolution is comparable or even finer than that in a previous numerical study \citep{han2021pdf} for the same Sandia pool fire.

All simulations are conducted using an in-house LES solver \citep{wen2021flamelet, wen2021flamelet2, wen2023flamelet, wen2023flamelet2, wen2024four}, which is developed based on OpenFOAM \citep{weller1998tensorial}. The LES solver has been comprehensively validated in previous works \citep{wen2021flamelet, wen2021flamelet2, wen2023flamelet, wen2023flamelet2, wen2024four}. The governing equations are solved with a finite volume method using the PIMPLE algorithm. The linear-upwind stabilized transport (LUST) scheme \citep{ControllingtheComputationalModesoftheArbitrarilyStructuredCGrid} is used to calculate the spatial integration of the momentum and scalar equations and the backward scheme is adopted to calculate the time integration. The time step is determined based on the Courant number, which is set at 1 to achieve stable simulations. Time averaging is started after the turbulent flow becomes statistically stable.

\section{Results and discussion}
\label{Sec:4}

\subsection{Comparison against experimental data \label{Subsec:41}}

\begin{figure}[!h]
    \centering
    \captionsetup[subfigure]{labelformat=empty} 
    \subfloat[]{
    \begin{minipage}[b]{0.55\textwidth}
\tikz[remember picture] \node[inner sep=0pt,outer sep=0pt] (a) 
{\includegraphics[trim = 0mm 0mm 0mm 0mm, clip, angle=0, width=1.\linewidth]{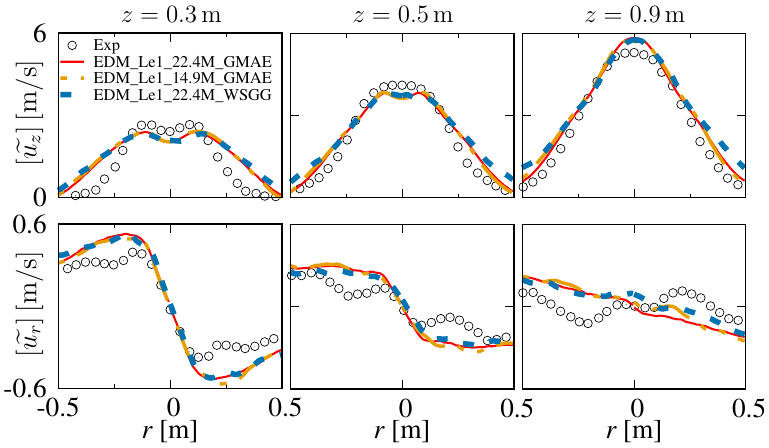}};
\end{minipage}
\begin{tikzpicture}[remember picture,overlay]
\end{tikzpicture}}
    \vspace{-5mm}
    \caption{\label{fig:velocity-mesh} Comparisons of the time-averaged axial and radial velocities between the experimental data and the simulation results obtained with different mesh resolutions (22.4\,M and 14.9\,M) and different gas absorption models (GMAE and WSGG) at three different axial locations. }
\end{figure}

First, the effects of the mesh resolution and radiation model on the prediction accuracy are evaluated by comparing the simulation results against the experimental data, as shown in Fig.~\ref{fig:velocity-mesh}. The evaluation is conducted based on the EDM model and the unity Lewis number assumption. The velocity profiles do not show obvious differences with different mesh resolutions and gas absorption models, which suggests that the mesh resolution is sufficiently fine and that radiative heat transfer does not affect the velocity distribution. Thus, the following discussions are based on the simulation results obtained with the finer mesh resolution (22.4\,M grid points) and the absorption coefficients calculated with the GMAE model.

\begin{figure}[!b]
    \centering
    \captionsetup[subfigure]{labelformat=empty} 
    \subfloat[]{
    \begin{minipage}[b]{0.55\textwidth}
\tikz[remember picture] \node[inner sep=0pt,outer sep=0pt] (a) 
{\includegraphics[trim = 0mm 0mm 0mm 0mm, clip, angle=0, width=1.\linewidth]{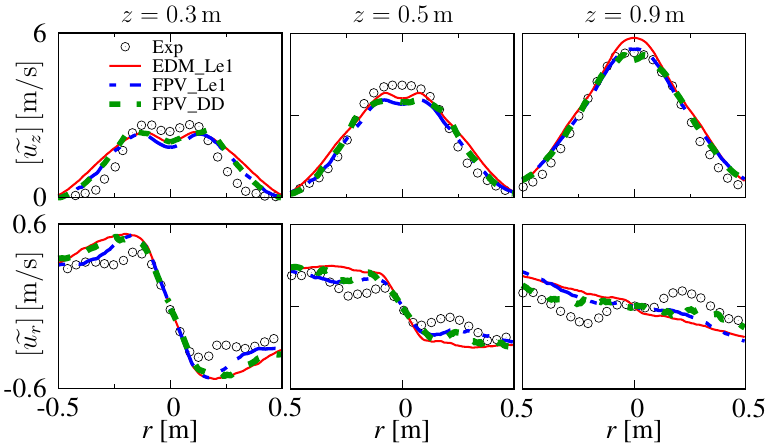}};
\end{minipage}
\begin{tikzpicture}[remember picture,overlay]
\end{tikzpicture}}
    \vspace{-5mm}
    \caption{\label{velocity} Comparisons of the time-averaged axial (upper row) and radial (lower row) velocities between the simulation obtained from different combustion models and the experimental data.  }
\end{figure}

To evaluate the suitability of the three combustion models, i.e., $\rm EDM\_Le1$, $\rm FPV\_Le1$ and $\rm FPV\_DD$, the time-averaged axial and radial velocities are compared between the simulation and the experiment, as shown in Fig.~\ref{velocity}. The simulation results obtained from the three combustion models agree reasonably well with the experiment. In most of the locations, the $\rm FPV\_Le1$ model gives better predictions compared to the $\rm EDM\_Le1$ model based on the infinitely fast chemistry assumption. Comparing the profiles of $\rm FPV\_Le1$ and $\rm FPV\_DD$ indicates that the inclusion of differential diffusion does not influence the prediction accuracy of the velocities. \textcolor{black}{This is because differential diffusion exerts a negligible effect on large-scale turbulent structures, consistent with the findings reported in the previous works for turbulent combustion simulations, e.g., see ref.~\cite{liu2026differential}.} The temporal vortical structures of the fire plume are analyzed and validated against available experimental data, and these validations are provided in the Supplementary Material.

\subsection{Overall combustion characteristics \label{Subsec:42}}
\begin{figure}[!h]
    \centering
    \captionsetup[subfigure]{labelformat=empty} 
    \subfloat[]{
    \begin{minipage}[b]{0.55\textwidth}
\tikz[remember picture] \node[inner sep=0pt,outer sep=0pt] (a) 
{\includegraphics[trim = 0mm 0mm 0mm 0mm, clip, angle=0, width=1.\linewidth]{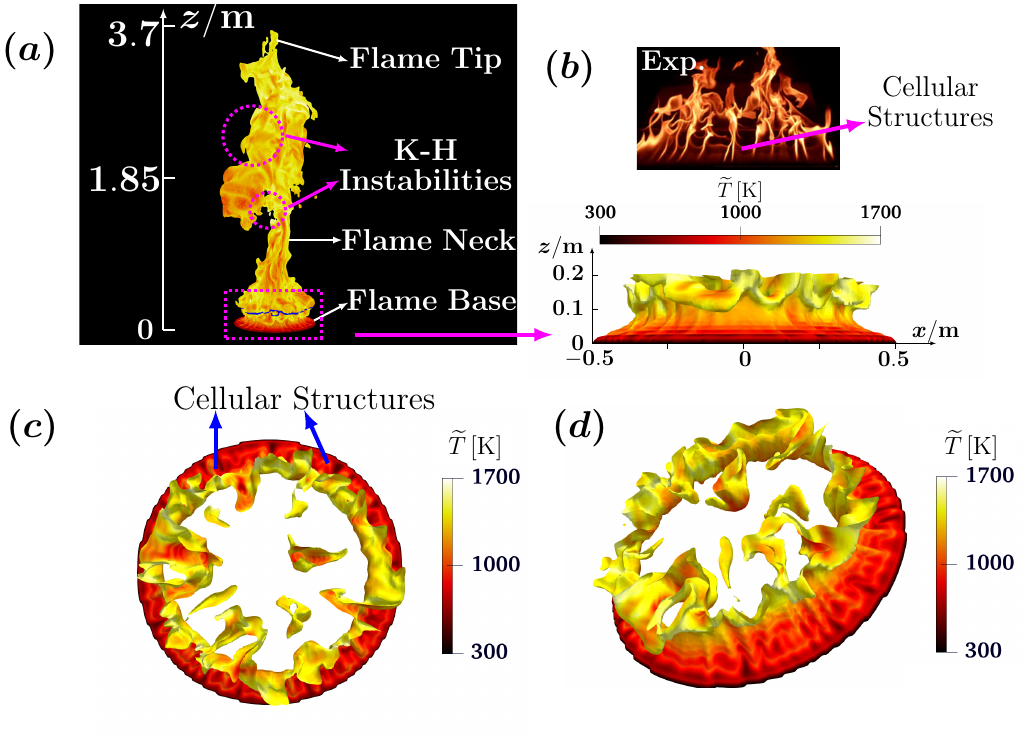}};
\end{minipage}
\begin{tikzpicture}[remember picture,overlay]
\end{tikzpicture}}
    \vspace{-9mm}
    \caption{\label{fig:flame} The 3D iso-surface of the stoichiometric mixture fraction $Z_{st}$ obtained with the $\rm FPV\_DD$ model colored by the local temperature value. (a) An overview of the $Z_{st}$ iso-surface, where the flame neck, flame base, and flame tip are indicated. The Kelvin-Helmholtz (K-H) instabilities are also indicated. (b) The further zoomed-in region as indicated in (a), and the experimental image of a wind-blown flame with coherent streaks from Ref.~\citep{gorham2014studying} is superimposed. (c) Top view of the same flame structure in (b), in which the cellular structures are indicated. (d) An overview of the same flame structure in (b) and (c). }
\end{figure}

An overview of the 3D iso-surface of the stoichiometric mixture fraction $Z_{st}$ obtained with the $\rm FPV\_DD$ model is shown in Fig.~\ref{fig:flame}a, and the flame front is colored by the local temperature value. Three different parts of the pool fire, including the flame base, flame neck, and flame tip, are indicated in Fig.~\ref{fig:flame}a, which represents the oscillating characteristics in a typical gaseous pool fire. The Kelvin-Helmholtz (K-H) instabilities are also indicated. A further zoomed-in region in $z < 0.2$\,m is presented in \ref{fig:flame}b, the top view of the same flame structure is given in Fig.~\ref{fig:flame}c, and the entire near-flow field is highlighted in Fig.~\ref{fig:flame}d. As indicated in Figs.~\ref{fig:flame}b and \ref{fig:flame}c, cellular structures can be observed close to the flame base with positively- and negatively-curved regions appearing periodically in the circumferential direction. Similar cellular structures were observed in the experiments, e.g., see the superimposed wind-blown flame with coherent streaks in Ref.~\citep{gorham2014studying}, and in the simulations for the pool fire \citep{taha2024large, domino2021predicting}. \textcolor{black}{Note that such cellular flame structures can also be observed in the gaseous pool fire without differential diffusion being considered (not shown), i.e., the $\rm {FPV\_Le1}$ and $\rm {EDM\_Le1}$ cases, which indicates that the cellular structure at the flame base does not result from thermodiffusive instabilities as in fuel-lean hydrogen flames with effective small Lewis numbers \cite{wen2022flamepart1, wen2024thermodiffusivelypart1}.} The mechanism for the formation of the cellular structure at the flame base is investigated in the next subsection. 

\subsection{Combustion instabilities}
\label{Subsec:43} %

Considering that the combustion instabilities at the flame base are governed by the vorticity equation, the contributions of different processes can be quantified according to the vorticity equation \cite{taha2024large}:
\begin{equation}\label{vorticity}
\begin{aligned}
\frac{D\widetilde{\mathcal{E}}}{Dt} &= \underbrace{(\widetilde{\mathcal{E}} \cdot \nabla)\widetilde{\bm{u}}}_{\text{vortex stretching}} 
- \underbrace{\widetilde{\mathcal{E}}(\nabla \cdot \widetilde{\bm{u}})}_{\text{dilatation term}} 
+ \underbrace{\frac{1}{\bar{\rho}^2}(\nabla\bar{\rho} \times \nabla \bar{p})}_{\text{baroclinic torque}} \\
&\quad + \underbrace{\frac{\rho_{\infty}}{\bar{\rho}^2}(\nabla\bar{\rho} \times \bm{g})}_{\text{gravitational torque}} 
+ \underbrace{\nabla \times \left(\frac{1}{\bar{\rho}}\nabla \cdot \tau\right)}_{\text{viscous diffusion}}
\end{aligned}
\end{equation}
where $\mathcal{E}$ is vorticity, $\rho_{\infty}$ the density of air, $\tau$ stress tensor. The right-hand side of Eq.~\eqref{vorticity} comprises five distinct terms representing different physical mechanisms governing vorticity transport: vortex stretching, dilatation, baroclinic torque, gravitational torque, and viscous diffusion. Vortex stretching characterizes the enhancement of vorticity by stretching and plays a fundamental role in the energy transfer process within turbulent flows. The dilatation term characterizes the weakening of vorticity due to volumetric expansion of fluid elements. The baroclinic torque quantifies the production of vorticity due to the misalignment between pressure and density gradients, while gravitational torque generates vorticity when density gradients are misaligned with the gravity vector. The viscous diffusion term represents the spatial diffusion and molecular dissipation of vorticity.

\begin{figure}[!h]
    \centering
    \captionsetup[subfigure]{labelformat=empty} 
    \subfloat[]{
    \begin{minipage}[b]{0.4\textwidth}
\tikz[remember picture] \node[inner sep=0pt,outer sep=0pt] (a) 
{\includegraphics[trim = 0mm 0mm 0mm 0mm, clip, angle=0, width=1.\linewidth]{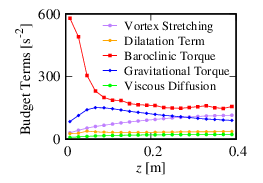}};
\end{minipage}
\begin{tikzpicture}[remember picture,overlay]
\end{tikzpicture}}
    \vspace{-6mm}
    \caption{\label{fig:vorticity} Comparisons of the time-averaged budget terms in the vorticity equation along the axial direction close to the flame base obtained with the $\rm FPV\_DD$ model.}
\end{figure}

To investigate the contribution of each budget term to vorticity transport, the time-averaged vorticity budget terms in the vorticity equation are presented in Fig.~\ref{fig:vorticity} for the $\rm FPV\_DD$ case. The baroclinic torque dominates over the other terms, exhibiting a peak near the inlet where the cellular structures are formed as visualized in Fig.~\ref{fig:flame}b. This indicates that cellular structures are predominantly generated by baroclinic torque resulting from the misalignment between pressure and density gradients. 



\begin{figure}[!h]
    \centering
    \captionsetup[subfigure]{labelformat=empty} 
    \subfloat[]{
    \begin{minipage}[b]{0.5\textwidth}
\tikz[remember picture] \node[inner sep=0pt,outer sep=0pt] (a) 
{\includegraphics[trim = 0mm 0mm 0mm 0mm, clip, angle=0, width=1.\linewidth]{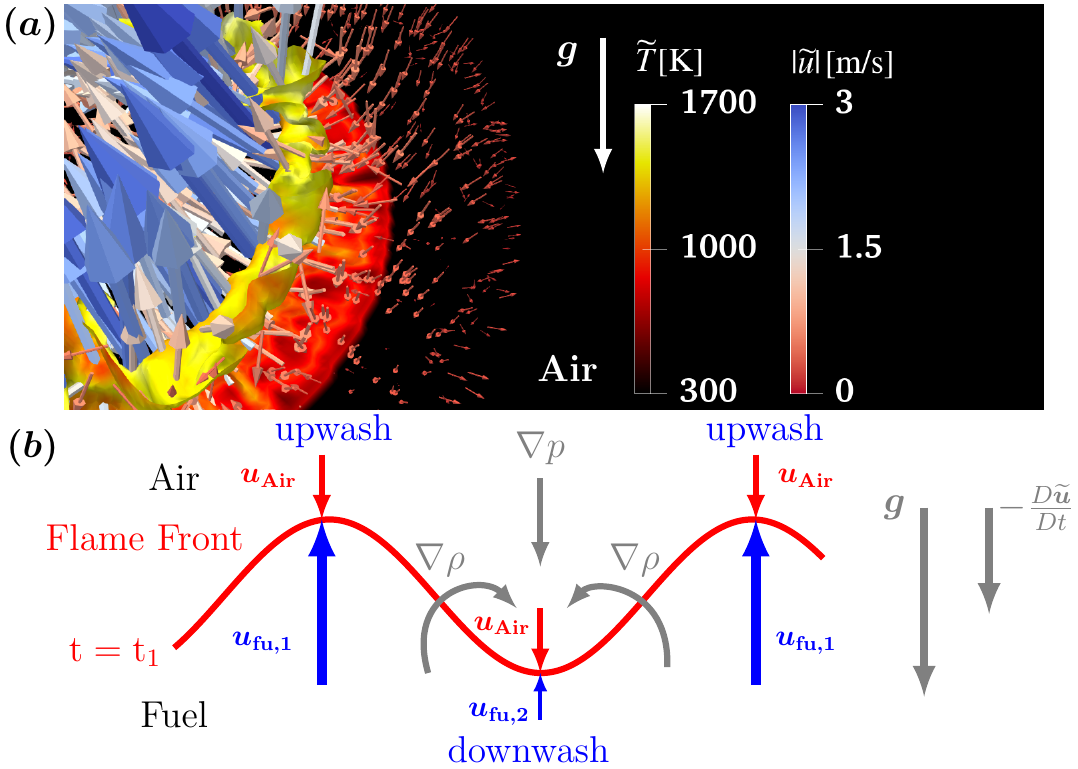}};
\end{minipage}
\begin{tikzpicture}[remember picture,overlay]
\end{tikzpicture}}
    \vspace{-5mm}
    \caption{\label{fig:Zst-black} (a) The 3D iso-surface of the stoichiometric mixture fraction $Z_{st}$ colored by the local temperature value close to the fuel side obtained with the $\rm FPV\_DD$ model. The velocity vectors colored by their magnitudes are superimposed to highlight the flow-flame structure interaction. (b) Schematic figure illustrating the cellular structure generation in the flame base.}
\end{figure}

The Rayleigh-Taylor (R-T) instability is the result of baroclinic torque created by the misalignment of the pressure and density gradients. To visualize the contribution from the Rayleigh-Taylor instability to the cellular structures, Fig.~\ref{fig:Zst-black}a presents a zoomed-in region of the 3D iso-surface of the stoichiometric mixture fraction $Z_{st}$ close to the fuel source for the $\rm FPV\_DD$ case, with the velocity vectors colored by their magnitudes superimposed to illustrate the local flow-flame interaction. The surrounding flow is entrained into the combustion region due to the buoyancy effects. The density differences generate a pair of counter-rotating streamwise vortices, which interact with the flame front. While the velocity magnitudes on the air side ${u_{\mathrm{air}}}$ are small and similar at different locations, they are significantly larger and different on the fuel side ${u_{fu}}$ due to the counter-rotating vortices. The resulting velocity difference lifts the flame in the upwash region while suppressing it in the downwash region, exacerbating the development of cellular structures at the flame base and further intensifying the misalignment of pressure and density gradients, which in turn strengthens the baroclinic torque instability. The R-T instability generates larger vortex structures in the downstream region, which further develop into Kelvin-Helmholtz (K-H) instability, as indicated in Fig. \ref{fig:flame}a. While the cellular structures close to the flame base are associated with the R-T instability generated by baroclinic torque, the combustion instabilities downstream, such as the puffing type instability, are mainly attributed to the K-H instability.

\subsection{Analyses of finite rate chemistry and differential diffusion}
\label{Subsec:44} %

\begin{figure}[!h]
    \centering
    \captionsetup[subfigure]{labelformat=empty} 
    \subfloat[]{
    \begin{minipage}[b]{0.55\textwidth}
\tikz[remember picture] \node[inner sep=0pt,outer sep=0pt] (a) 
{\includegraphics[trim = 0mm 0mm 0mm 0mm, clip, angle=0, width=1.\linewidth]{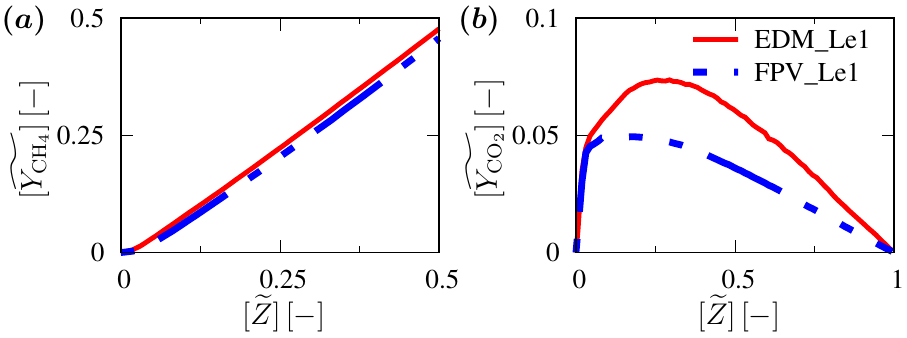}};
\end{minipage}
\begin{tikzpicture}[remember picture,overlay]
\end{tikzpicture}}
    \vspace{-5mm}
    \caption{\label{fig:EDM-Le1} Comparisons of the time-averaged (a) \ce{CH4} mass fraction and (b) \ce{CO2} mass fraction in the mixture fraction space obtained from the $\rm EDM\_Le1$ and $\rm FPV\_Le1$ models.}
\end{figure}

To quantify the effects of \textbf{finite rate chemistry} on the prediction of the thermo-chemical quantities, the time-averaged \ce{CH4} and \ce{CO2}  mass fractions predicted by the $\rm EDM\_Le1$ and $\rm FPV\_Le1$ models are shown in Fig.~\ref{fig:EDM-Le1}. The \ce{CH4} mass fraction obtained with the $\rm EDM\_Le1$ and $\rm FPV\_Le1$ models does not show obvious differences, see Fig.~\ref{fig:EDM-Le1}a, which suggests that the prediction of \ce{CH4} mass fraction is not sensitive to the chemistry assumption. However, the peak value of the \ce{CO2} mass fraction predicted with the $\rm EDM\_Le1$ model is much larger than that predicted with the $\rm FPV\_Le1$ model, as shown in Fig.~\ref{fig:EDM-Le1}b, which is associated with the fact that intermediate species such as CO are not considered in the single-step reaction mechanism.

\begin{figure}[!b]
    \centering
    \captionsetup[subfigure]{labelformat=empty} 
    \subfloat[]{
    \begin{minipage}[b]{0.55\textwidth}
\tikz[remember picture] \node[inner sep=0pt,outer sep=0pt] (a) 
{\includegraphics[trim = 0mm 0mm 0mm 0mm, clip, angle=0, width=1.\linewidth]{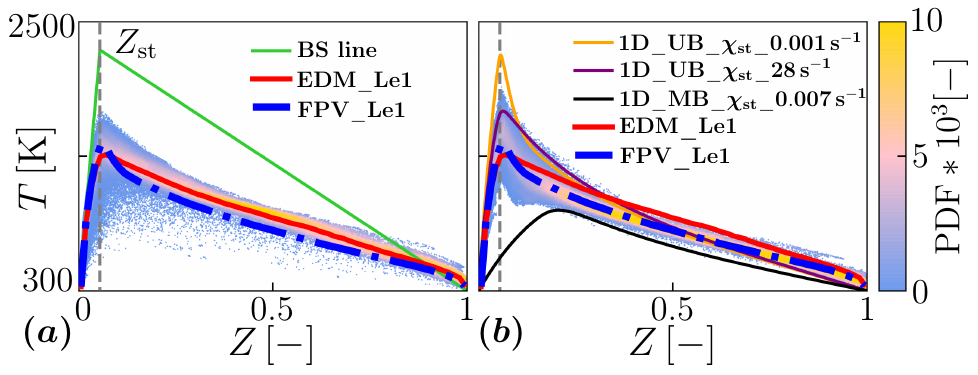}};
\end{minipage}
\begin{tikzpicture}[remember picture,overlay]
\end{tikzpicture}}
    \vspace{-5mm}
    \caption{\label{fig:EDM-Le1-T} Comparisons of the conditioned mean temperature between the $\rm EDM\_Le1$ and $\rm FPV\_Le1$ models. Scatter points are colored by PDF. The Burke–Schumann (BS) line is indicated in (a), while the flamelets evaluated at the maximum and minimum scalar dissipation rates in the upper branch (UB) of the S-shaped curve \cite{peters1984laminar}, and the minimum scalar dissipation rate in the middle branch (MB) are superimposed for comparison. The vertical line indicates the location of the stoichiometric mixture fraction $Z_{\rm st}$.}
\end{figure}

Figure \ref{fig:EDM-Le1-T} compares the conditional temperatures between the $\rm EDM\_Le1$ and $\rm FPV\_Le1$ models in the mixture fraction space. Scatter points, which are colored by PDF, are extracted from different time instants. The Burke–Schumann (BS) line in Fig.~\ref{fig:EDM-Le1-T}a shows the limit of infinitely fast irreversible chemistry, The peak value of the BS line corresponds to the adiabatic flame temperature, which is calculated as,
\begin{equation}\label{eq:Tad}
T_{\rm {ad}} = Z_{\rm {st}}T_{\rm F}^0 + (1-Z_{\rm {st}})T_{\rm O}^0 +\frac{QY_{\rm F}^0}{C_{\rm p}} 
\end{equation}
where $T_{\rm F}^0$ and $T_{\rm O}^0$ are the inlet temperatures of the fuel and air, respectively; $Q$ the specific heat of reaction under stoichiometric conditions, $C_{p}$ the specific heat capacity of the mixture, and $Y_{\rm F}^0$ the inlet fuel mass fraction. The calculated adiabatic temperature is $T_{\rm ad} = 2268\,\rm K$. The flamelets calculated at the maximum and minimum scalar dissipation rates in the upper branch (UP) of the S-shaped curve \cite{peters1984laminar} and the minimum scalar dissipation rate in the middle branch (MB) of the S-shaped curve are superimposed for comparison. The vertical line indicates the location of the stoichiometric mixture fraction $Z_{\rm {st}}$. The temperature predicted by the $\rm EDM\_Le1$ model is below the BS line, as expected. The conditioned mean temperature predicted by the $\rm EDM\_Le1$ model is slightly higher than that obtained with the flamelet model in the range of $Z>Z_{\rm st}$. As shown in Fig.~\ref{fig:EDM-Le1-T}b, the scatters obtained with the flamelet model are mostly located within the middle branch of the S-shaped curve, which indicates that transient states such as local extinction are prevalent in the large-scale pool fires studied.

\begin{figure}[!h]
    \centering
    \captionsetup[subfigure]{labelformat=empty} 
    \subfloat[]{
    \begin{minipage}[b]{0.55\textwidth}
\tikz[remember picture] \node[inner sep=0pt,outer sep=0pt] (a) 
{\includegraphics[trim = 0mm 0mm 0mm 0mm, clip, angle=0, width=1.\linewidth]{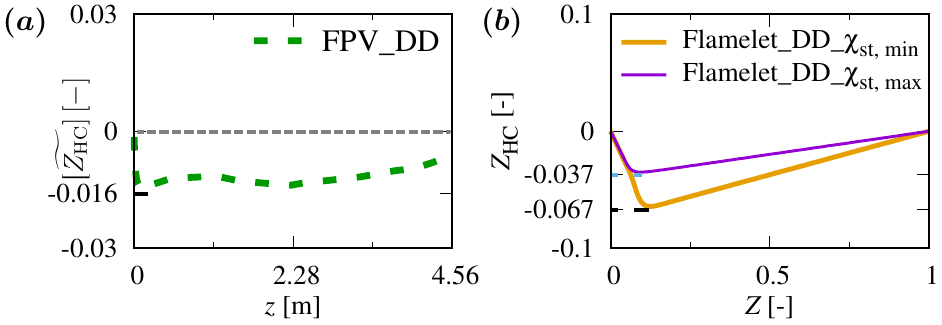}};
\end{minipage}
\begin{tikzpicture}[remember picture,overlay]
\end{tikzpicture}}
    \vspace{-5mm}
    \caption{\label{fig:ZHC} Comparisons of the time-averaged (a) \ce{CH4} mass fraction and (b) \ce{CO2} mass fraction in the mixture fraction space obtained from the $\rm EDM\_Le1$ and $\rm FPV\_Le1$ models.}
\end{figure}

\begin{figure}[!b]
    \centering
    \captionsetup[subfigure]{labelformat=empty} 
    \subfloat[]{
    \begin{minipage}[b]{0.55\textwidth}
\tikz[remember picture] \node[inner sep=0pt,outer sep=0pt] (a) 
{\includegraphics[trim = 0mm 0mm 0mm 0mm, clip, angle=0, width=1.\linewidth]{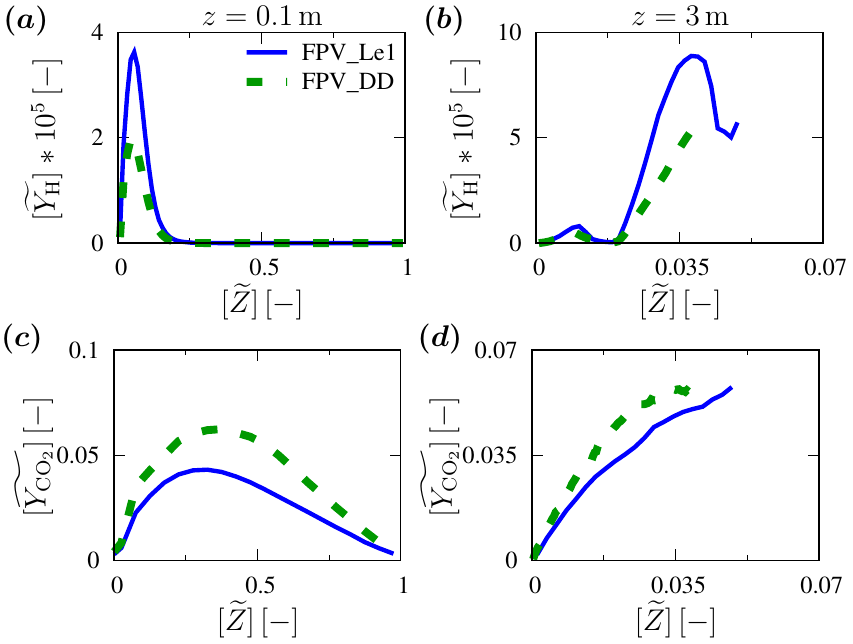}};
\end{minipage}
\begin{tikzpicture}[remember picture,overlay]
\end{tikzpicture}}
    \vspace{-5mm}
    \caption{\label{fig:DD-Le1} Comparisons of the time-averaged mass fractions of (a) H radical, and (b) \ce{CO2} in the mixture fraction space obtained with the $\rm FPV\_Le1$ and $\rm FPV\_DD$ models upstream at $z = 0.1\,\rm m$ (left column) and downstream at $z = 3\,\rm m$ (right column). }
\end{figure}

The importance of \textbf{differential diffusion} can be quantified by introducing the differential diffusion parameter. There could be multiple definitions for the differential diffusion parameter, such as $Z_{\rm HC}$, $Z_{\rm HO}$ and $Z_{\rm CO}$. For hydrocarbon combustion, $Z_{\rm HC}$ is generally used as a universal parameter to quantify the differential diffusion effect as most of the species contain H and C atoms. Following Barlow et al.~\citep{barlow2000experiments}, the differential diffusion parameter $Z_{\rm HC}$ is defined as the difference between the normalized elemental mixture fractions of H and C, i.e., 
\begin{equation}
Z_{\rm HC} = \xi_{\rm H} - \xi_{\rm C}
\end{equation}
where $ \xi_{\rm H}$ and $\xi_{\rm C}$ are calculated as,
\begin{equation}
\xi_{\rm H} = \frac{Z_{\rm H} - Z_{\mathrm{H, ox}}}{ Z_{\mathrm{H, fu}} - Z_{\mathrm{H, ox}} }
\end{equation}
\begin{equation}
\xi_{\rm C} = \frac{Z_{\rm C} - Z_{\mathrm{C, ox}}}{ Z_{\mathrm{C, fu}} - Z_{\mathrm{C, ox}} }
\end{equation}
where $Z_{\zeta}$ is the mixture fraction of element $\zeta$, and the subscripts $fu$ and $ox$ denote the fuel and oxidizer, respectively. The importance of differential diffusion can be quantified by the magnitude of $Z_{\mathrm{HC}}$ deviating from zero. The time-averaged distribution of $\widetilde{Z_{\rm HC}}$ obtained with  the $\rm FPV\_DD$ model is shown in Fig.~\ref{fig:ZHC}a along the streamwise direction of the fire plume. The significance of differential diffusion increases close to the fuel source, and then remains significant in the downstream flow regions, which confirms the significance of differential diffusion in the fire plume studied. The importance of differential diffusion in the whole region of the fire plume could be associated with the existence of puffing cycles, which results in the local flame neck with weak turbulence mixing. Note that the limit value of $\widetilde{Z_{\rm HC}}$ is approximately 0.016 in the LES, which is much smaller than the corresponding values calculated with the flamelets at the maximum and minimum scalar dissipation rates with differential diffusion, as shown in Fig.~\ref{fig:ZHC}b. This indicates that the effects of differential diffusion in the turbulent fire plume are less significant than those in the corresponding laminar flames. 


To quantify the effects of differential diffusion on the distributions of thermo-chemical quantities in the upstream (left column) and downstream (right column) regions, the time-averaged mass fractions of H radical and \ce{CO2} are shown in the mixture fraction space, see Fig.~\ref{fig:DD-Le1}. The H radical and \ce{CO2} are selected to represent the light and heavy species, respectively, with potential differential diffusion effects. It can be observed that in both upstream and downstream regions, the $\rm FPV\_DD$ model predicts a generally lower \ce{H} radical mass fraction and a higher \ce{CO2} concentration compared to the $\rm FPV\_Le1$ model. This demonstrates that differential diffusion significantly affects the distributions of light and heavy species in the whole domain, aligning with the distribution of the differential diffusion parameter shown in Fig.~\ref{fig:ZHC}a. Note that although the overall fire dynamics may not be affected by radicals, these radicals directly change the soot formation process. For example, the well-known H-abstraction-\ce{C2H2}-addition (HACA) process \cite{frenklach1991detailed} describes soot particle surface growth in a mathematically rigorous way, and the inaccurate prediction of H radical directly influences the HACA process, and thus the prediction of soot formation. Note that the peak value of the mixture fraction in the downstream region is much smaller due to the fuel consumption. In addition, the maximum mixture fraction of the $\rm FPV\_DD$ case is smaller than that of $\rm FPV\_Le1$, which results from the faster consumption rate with differential diffusion.

\section{Conclusion}

Large-eddy simulations are conducted for the Sandia one-meter diameter gaseous pool fire with the eddy dissipation model based on the unity Lewis number assumption ($\rm EDM\_Le1$), the flamelet models based on the unity Lewis number assumption ($\rm FPV\_Le1$) and the mixture-averaged approach to consider differential diffusion ($\rm FPV\_DD$). Comparing with the experiment, all combustion models give accurate predictions for the velocities, regardless of the treatments of differential diffusion and chemistry. Cellular structures are observed close to the fuel source with positively- and negatively-curved flame fronts appearing periodically. The mechanism for the formation of cellular structures is analyzed based on the vorticity equation. A \textit{budget analysis} verifies that the cellular structures at the flame base are associated with the Rayleigh-Taylor instability generated by baroclinic torque, while the Kelvin-Helmholtz instability associated with the puffing cycle dominates in the downstream region. The \ce{CO2} mass fraction is over-predicted by the $\rm EDM\_Le1$ model compared with the $\rm FPV\_Le1$ model due to the employed single-step reaction mechanism. In contrast, the temperature field is not significantly influenced by the finite-rate chemistry. The significance of differential diffusion increases close to the fuel source and remains important in the downstream flow regions. This subsequently affects the distributions of \ce{H} radical and \ce{CO2} mass fractions throughout the entire flow field, which are directly related to soot formation. Thus, it is necessary to incorporate differential diffusion in a flamelet model to improve the prediction accuracy.

\section*{CRediT authorship contribution statement}
Y.~Yan: Performed simulations, analyzed data, plotted figures, wrote paper.
F.~Brännström: Analyzed data, reviewing, editing, supervision.
C.~Hasse: Analyzed data, reviewing, editing, supervision. 
X.~Wen: Designed research, performed simulations, analyzed data, wrote paper, supervision.


\section*{Declaration of competing interest}

The authors declare that they have no known competing financial interests or personal relationships that could have appeared to influence the work reported in this paper.

\section*{Acknowledgments}

X. Wen acknowledges the financial support from the National Natural Science Foundation of China for the Excellent Young Scientists Fund Program.

\section*{Supplementary material}

Supplementary Material is provided to support the main text of this paper.

\section*{Data availability}

Data will be made available on request.



\bibliographystyle{elsarticle-num}
\bibliography{sci}




\end{document}

